\documentclass[prl,twocolumn,floatfix,
amsmath,amssymb,aps,superscriptaddress,nofootinbib]{revtex4-1}
\usepackage{dcolumn,amsmath}
\usepackage[T2A]{fontenc}
\usepackage{graphicx}
\graphicspath{ {./Images/} }
\usepackage{bm}
\usepackage{xfrac}
\usepackage{indentfirst} 
\usepackage{physics}

\usepackage{xcolor}

\usepackage{hyperref}

\newcommand{\LS}[1]{\textcolor{blue}{#1}}

\usepackage{amsmath}
\usepackage{graphicx}
\usepackage{bm}
\usepackage{xcolor}
\usepackage{multirow}
\usepackage{amssymb}
\usepackage{soul}

\newcommand{\eEDM}{{\em e}EDM}
\newcommand{\ecm}{\ensuremath{e {\cdotp} {\rm cm}}}

\newcommand{\de}{d_\mathrm{e}}

\newcommand{\tref}[1]{Table~\ref{#1}}

\newcommand{\B}{\mathcal{B}} 

\newcommand{\Brot}{B_\mathrm{rot}}

\begin{document}
\title{Electric field dependent g factors of ThF$^+$}

\begin{abstract}
The g-factors for $J = 1$, $F=3/2$, $|M_F|=3/2$ hyperfine levels of the ground electronic state $^3\Delta_1$ of the $^{232}$ThF$^+$ cation are calculated as functions of the external electric field. These calculations are necessary for the analysis of systematic effects in the experiment aimed at searching for the electron electric dipole moment.
\end{abstract}

\author{Alexander Petrov}
\email{petrov\_an@pnpi.nrcki.ru}
\author{Leonid V. Skripnikov}
\email{skripnikov_lv@pnpi.nrcki.ru}
\affiliation{Petersburg Nuclear Physics Institute named by B.P.\ Konstantinov of National Research Center ``Kurchatov Institute'' (NRC ``Kurchatov Institute'' - PNPI), 1 Orlova roscha mcr., Gatchina, 188300 Leningrad region, Russia}
\affiliation{Saint Petersburg State University, 7/9 Universitetskaya nab., St. Petersburg, 199034 Russia}
\homepage{http://www.qchem.pnpi.spb.ru    }


\maketitle

\section{Introduction}
The measurement of the electric dipole moment of the electron (\eEDM) serves as a highly sensitive probe to test the limits of the Standard Model of electroweak interactions and its extensions \cite{whitepaper, YamaguchiYamanaka2020,YamaguchiYamanaka2021,Safronova:18,arrowsmithkron2023opportunities,KL95}. The current constraint on \eEDM, $|d_e|<4.1\times 10^{-30}$~\ecm (90\% confidence level), was obtained using trapped $^{180}$Hf$^{19}$F$^+$ ions \cite{newlimit1} with the spinless $^{180}$Hf isotope. The measurements were performed on the ground rotational level, $J{=}1$, in the {\it excited} metastable electronic $^3\Delta_1$ state.
It improves the ACME collaboration result obtained in 2018, $|d_e| \lesssim 1.1\cdot 10^{-29}\ e\cdot\textrm{cm}$ \cite{ACME:18}, by a factor of 2.4 and the first result $|\de|\lesssim 1.3\times 10^{-28}$ on the $^{180}$Hf$^{19}$F$^+$ ions  \cite{Cornell:2017} by a factor of about 32.
Recently, it was proposed to conduct \eEDM~search experiments using the isovalent $^{232}$ThF$^+$ cation \cite{Gresh:2016, Ng:2022}, also with the spinless $^{232}$Th isotope. In ThF$^+$, the \eEDM-sensitive state, $^3\Delta_1$, is the electronic {\it ground} state, which allows for a longer interrogation time. Furthermore, improvements in the design of the ion trap enable a larger number of ions to be used for statistical analysis \cite{Ng:2022}. Finally, the effective electric field, $E_{\rm eff} = 37.3$~GV/cm \cite{Skripnikov:2015t,Fleig:15}, in ThF$^+$ is stronger than the corresponding field, $E_{\rm eff} = 22.5$~GV/cm \cite{Skripnikov:17c,Fleig:17, Petrov:07a}, in HfF$^+$. These factors collectively enhance the experimental sensitivity of ThF$^+$ to \eEDM~and other $\mathcal{T,P}$-odd effects ($\mathcal{P}$ -- spatial parity and $\mathcal{T}$ -- time reversal).

The \eEDM\ measurements using HfF$^+$ and ThF$^+$ are, in fact, highly accurate spectroscopy of the energy splitting between levels with opposite directions of the total
angular
momentum projections of the $J{=}1$ level in the presence of rotating electric and magnetic fields. Beyond the \eEDM, there is, for example, the Zeeman-induced contribution to the splitting. Thus, insufficient control of the magnetic field is a source of systematic errors in the experiment. The accurate evaluation of systematic effects becomes very important with the planned increase in statistical sensitivity \cite{Ng:2022}.

A major factor contributing to the great success in addressing this problem in the HfF$^+$ experiment is the existence of closely spaced levels, known as the $\Omega$-doublet, of opposite parities, which become a Stark doublet in the presence of an electric field. The measurement of the energy splitting can be performed on both levels of the Stark doublet. The key point is that the structure of the Stark doublet is arranged in such a way that the \eEDM\ contribution to the splitting is opposite in the two Stark doublet states, whereas the Zeeman contributions have the same sign. As 
noted in Ref. \cite{DeMille2001}, the energy splittings in the two Stark doublet states can be subtracted from each other, thereby suppressing systematic effects related to stray magnetic fields (as well as many other systematic effects) while doubling the \eEDM\ signal.

Unfortunately, the upper and lower states of the Stark doublet have slightly different magnetic g-factors, and the systematic error is not completely canceled. The difference $\Delta {\rm g}$ depends on the laboratory electric field. Therefore, it is clear that understanding the dependence of the g-factor on the electric field is crucial for controlling and suppressing possible systematic effects in \eEDM\ search experiments. Previously, the dependence of the g-factor on the electric field for HfF$^+$ was calculated in Refs \cite{Petrov:17b, Petrov:18, Kurchavov:2020, Kurchavov2021} 

Our method demonstrated very high accuracy \cite{Petrov:23, Petrov:23b}. In the current work, we calculated the g-factors of $^{232}$ThF$^+$ in the ground rotational level $J{=}1$ of the ground electronic state $^3\Delta_1$ as a function of the external electric field. Using the obtained values, we analyze systematic effects connected with magnetic effects.

\section {The electron electric dipole moment sensitive levels}

\begin{figure}
    \centering
    \includegraphics[width=1.0\linewidth]{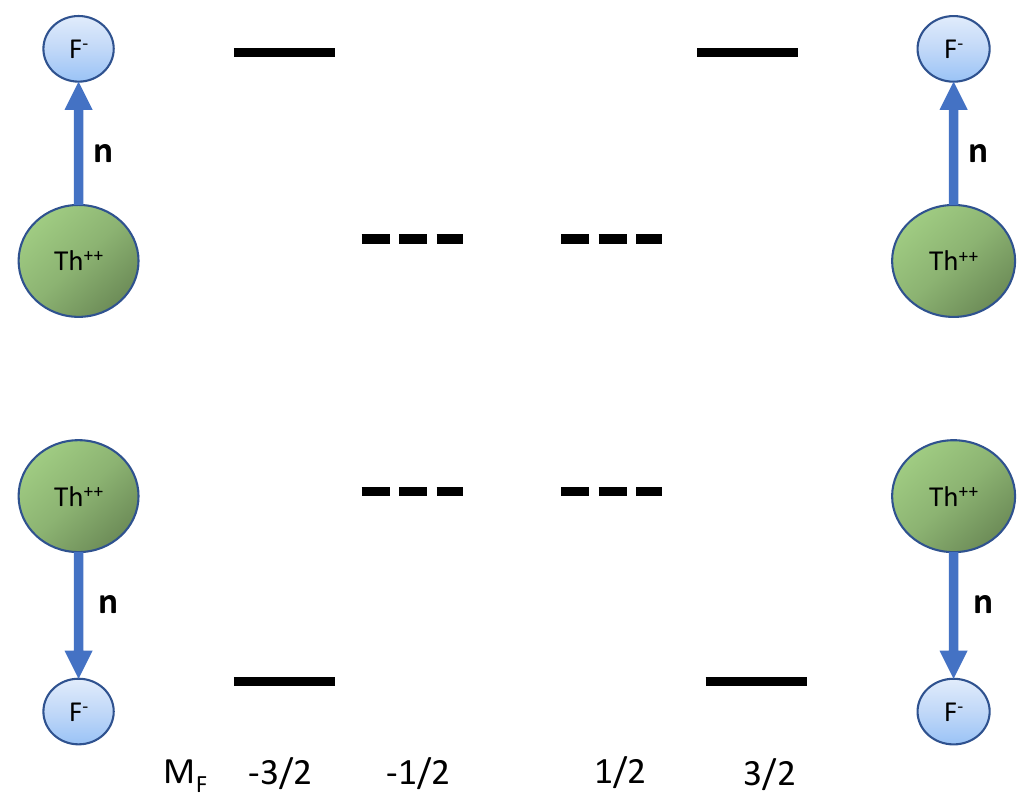}
    \caption{Energy levels of F=3/2 in the external electric field. Solid lines is the Stark doublet for $|M_F|=3/2$. Dashed lines is the Stark doublet for $|M_F|=1/2$. Stark doublet for $|M_F|=3/2$ is the focus of the \eEDM\ search experiment.}
    \label{Fig1}
\end{figure}

The $^{232}$Th isotope is spinless, whereas the $^{19}$F isotope has a nonzero nuclear spin, $I{=}1/2$. A hyperfine energy splitting between the levels with total momentum $F=3/2$ and $F=1/2$, where ${\bf F}={\bf J}+{\bf I}$, in the ground rotational level, $J=1$, is about 15~MHz \cite{Ng:2022}. In the absence of external fields, each hyperfine level has two parity eigenstates, with a splitting of about 5.3 MHz \cite{Ng:2022}, known as the $\Omega$-doublets. An external {\it static} electric field mixes levels of opposite parities such that the $F=3/2$ states ($F=1/2$ is not the subject of this work) form two Stark doublet levels, with absolute values of the projection of the total 
angular
momentum onto the direction of the electric field, $M_F$, equal to one-half and three-halves as shown in Fig.~\ref{Fig1}. 
Below, the levels of the Stark doublet with $|M_F|=3/2$ will be referred to as upper and lower according to their energies. The upper and lower levels in the doublet are doubly degenerate. Specifically, if \eEDM\ is zero, two Zeeman sublevels connected by time reversal, $M_F \rightarrow -M_F$, have the same energy. The levels $M_F=\pm3/2$ are of particular interest for the \eEDM\ search experiment. As stated above, their splitting, caused by the hypothetical \eEDM, is the primary focus of the experiment.

\section{Theoretical methods}
\subsection{Molecular calculation}
Following Refs.~\cite{Petrov:11, Petrov:17b, Petrov:18}, the energy levels and wave functions of the $^{232}$Th$^{19}$F$^+$ ion are obtained by numerical diagonalization of the molecular Hamiltonian (${\rm \bf \hat{H}}_{\rm mol}$) in the presence of external static electric ${\bf E}$ and magnetic ${\bf B}$ fields over the basis set of the electronic-rotational wavefunctions
\begin{equation}
 \Psi_{\Omega}\theta^{J}_{M,\Omega}(\alpha,\beta)U^{\rm F}_{M_I}.
\label{basis}
\end{equation}
Here, $\Psi_{\Omega}$ is the electronic wavefunction, $\theta^{J}_{M,\Omega}(\alpha,\beta)=\sqrt{(2J+1)/{4\pi}}D^{J}_{M,\Omega}(\alpha,\beta,\gamma=0)$ is the rotational wavefunction, where $\alpha,\beta,\gamma$ are Euler angles. Additionally, $U^{F}_{M_I}$ represents the nuclear spin wavefunction, and $M$ $(\Omega)$ denotes the projection of the molecular angular momentum, ${\bf J}$, onto the laboratory $\hat{z}$ (internuclear $\hat{n}$) axis. The nuclear 
spin
projection is given by $M_I=\pm1/2$, and the total momentum projection is $M_F=M_I+M$.

The molecular Hamiltonian for 
$^{232}$ThF$^+$
reads
\begin{equation}
{\rm \bf\hat{H}}_{\rm mol} = {\rm \bf \hat{H}}_{\rm el} + {\rm \bf\hat{H}}_{\rm rot} + {\rm \bf\hat{H}}_{\rm hfs} + {\rm \bf\hat{H}}_{\rm ext}.
\end{equation} 
Here, ${\rm \bf \hat{H}}_{\rm el}$ is the electronic Hamiltonian, ${\rm \bf\hat{H}}_{\rm rot}$ is the Hamiltonian describing the rotation of the molecule, ${\rm \bf\hat{H}}_{\rm hfs}$ represents the hyperfine interaction between electrons and fluorine nuclei, and ${\rm \bf\hat{H}}_{\rm ext}$ describes the interaction of the molecule with external magnetic and electric fields, as detailed in Ref. \cite{Petrov:17b}.

In the present work, we considered the low-lying electronic basis states $^3\Delta_1$, $^1\Sigma_0^+$, $^3\Pi_{0^+}$, $^3\Pi_{0^-}$, and $1,2,3^3\Delta_2$. The electronic Hamiltonian ${\rm \bf \hat{H}}_{\rm el}$ is diagonal in the basis set (\ref{Molbasis}), and its eigenvalues correspond to the excitation energies of these states.

\begin{align}
\label{Molbasis}
\nonumber
^3\Delta_1 & : T_0=0~{\rm cm}^{-1}\ ,\\
\nonumber
 ^1\Sigma^+ & : T_0=314~{\rm cm}^{-1}\ ,\\
\nonumber
 1^3\Delta_2 & : T_0=1052~{\rm cm}^{-1}\ ,\\
 \nonumber
 2\Omega=2 & : T_v=5806~{\rm cm}^{-1}\ ,\\
 \nonumber
 3\Omega=2 & : T_v=6402~{\rm cm}^{-1}\ ,\\
 \nonumber
^3\Pi_{0^-} & : T_0=3044~{\rm cm}^{-1}\ ,\\
 ^3\Pi_{0^+} & : T_0=3395~{\rm cm}^{-1}\ .
\end{align}
  
The $T_0$ values for $^1\Sigma^+$, $1^3\Delta_2$, and $^3\Pi_{0^+}$ were taken from Ref. \cite{Barker:2012}, while the values for other states were calculated in the present work as a vertical transition ($T_v$) at a near equilibrium internuclear distance $R=3.75$~Bohr.

The electronic matrix elements used to calculate the matrix element of the molecular Hamiltonian, $ {\rm \bf \hat{H}}_{\rm mol}$, between different electronic states are given by Eqs. (\ref{Gperp1})--(\ref{Delta6}) below.

We have performed electronic calculations for the internuclear distance R=3.75 Bohr for the following matrix elements of the basis electronic states:

\begin{eqnarray}
 \label{Gperp1}
   G_{\perp}^{(1)} &=& \langle  ^3\Delta_1  |\hat{L}^e_+ -  {\rm g}_{S}\hat{S}^e_+ |^1\Sigma^+  \rangle = 0.48, \\
   G_{\perp}^{(2a)} &=& \langle  1^3\Delta_2  |\hat{L}^e_+ -  {\rm g}_{S}\hat{S}^e_+ |^3\Delta_1  \rangle = -2.41, \\
   G_{\perp}^{(2b)} &=& \langle  2\Omega=2  |\hat{L}^e_+ -  {\rm g}_{S}\hat{S}^e_+ |^3\Delta_1  \rangle = -0.06, \\
   G_{\perp}^{(2c)} &=& \langle  3\Omega=2  |\hat{L}^e_+ -  {\rm g}_{S}\hat{S}^e_+ |^3\Delta_1  \rangle = 0.93, \\
   G_{\perp}^{(3a)} &=& \langle  ^3\Delta_1  |\hat{L}^e_+ -  {\rm g}_{S}\hat{S}^e_+ |^3\Pi_{0^-}  \rangle = 1.07, \\
   G_{\perp}^{(3b)} &=& \langle  ^3\Delta_1  |\hat{L}^e_+ -  {\rm g}_{S}\hat{S}^e_+ | ^3\Pi_{0^+} \rangle = 1.01, 
\end{eqnarray} 

\begin{eqnarray}
 \label{Delta1}
  \Delta^{(1)} &=& 2\Brot\langle  ^3\Delta_1  |\hat{J}^e_+  |^1\Sigma^+  \rangle = 0.250~{\rm cm}^{-1}, \\
   \Delta^{(2a)} &=& 2\Brot\langle  1^3\Delta_2  |\hat{J}^e_+  |^3\Delta_1  \rangle = -0.510~{\rm cm}^{-1}, \\
   \Delta^{(2b)} &=& 2\Brot\langle  2\Omega=2  |\hat{J}^e_+  |^3\Delta_1  \rangle = -0.113~{\rm cm}^{-1}, \\
   \Delta^{(2c)} &=& 2\Brot\langle  3\Omega=2  |\hat{J}^e_+  |^3\Delta_1  \rangle = 0.427~{\rm cm}^{-1}, \\
   \Delta^{(3a)} &=& 2\Brot\langle  ^3\Delta_1  |\hat{J}^e_+  |^3\Pi_{0^-}  \rangle = 0.587~{\rm cm}^{-1}, \\
\label{Delta6}
   \Delta^{(3b)} &=& 2\Brot\langle  ^3\Delta_1  |\hat{J}^e_+  | ^3\Pi_{0^+} \rangle = 0.555~{\rm cm}^{-1}, 
\end{eqnarray} 
where $\Brot=0.243~{\rm cm}^{-1}$ \cite{Ng:2022} is the rotational constant, ${\rm g}_{S} = -2.0023$ is the free-electron $g$-factor, $\vb{J}^e = \vb{L}^e + \vb{S}^e$ represents the total, orbital, and spin angular momenta of the electronic subsystem, $J^e_{\pm} = J^e_{x} \pm iJ^e_{y}$, and the same applies for other vectors.
We found that matrix element (\ref{Gperp1}) provides the main contribution to the $\Omega$-doubling effect. It was adjusted to reproduce the experimental value of 5.29 MHz for $\Omega$-doubling measured in Ref. \cite{Ng:2022}. To achieve this, we increased the calculated value by 8.8\%. Due to the similarity between matrix elements (\ref{Gperp1}) and (\ref{Delta1}), matrix element (\ref{Delta1}) was also increased by 8.8\%. For other matrix elements (\ref{Gperp1})--(\ref{Delta6}), we used {\it ab initio} values calculated in the present work. We used $D=-0.133$ a.u. (\LS{with respect to center of nuclear mass;} the molecular axis directed from Th to F) and $A_{\parallel}=-20.1$ MHz from Ref. \cite{Ng:2022}.
We also used $G_{\parallel}=0.047$ to reproduce the ${\rm g}=0.0149$ value for the g-factor of the $F=3/2$ state. Our value is slightly different from $G_{\parallel}=0.048$ estimated in Ref. \cite{Ng:2022} since we account for nonadiabatic interactions between $^3\Delta_1$ and other states in the basis set (\ref{Molbasis}).

Beyond those listed in Eq. (\ref{Molbasis}), there are other electronic states that can interact with $^3\Delta_1$ (the $e$EDM-sensitive state). However, states with $\Omega > 2$ are not mixed at leading order due to selection rules and can be safely ignored. The $^3\Pi_{0^-}$ and $^3\Pi_{0^+}$ states in Eq. (\ref{Molbasis}) have the same configuration ($\sim$7s$^1$6d$^1$, where 7s and 6d are atomic orbitals of Th) as $^3\Delta_1$, and therefore, they have the largest matrix elements among the $\Omega=0^{\pm}$ states. 
The $^1\Sigma^+$ state is very close in energy to $^3\Delta_1$ and must be taken into account. Matrix elements between $^1\Sigma^+$ and $^3\Delta_1$ were modified to fit the experimental value of the $\Omega$-doubling, which, as first noted in Ref. \cite{Petrov:17b}, effectively accounts for interactions with other $\Omega=0$ states.
Among the $\Omega=2$ states, the $1^3\Delta_2$ state from Eq. (\ref{Molbasis}) also has the largest matrix element with $^3\Delta_1$ (as they share the same configuration), and the energy difference is only about a thousand wave numbers. Other $\Omega=2$ states interact much more weakly. Nevertheless, we also included two more $\Omega=2$ states in our calculations.
Finally, we believe that the accuracy of the model is not worse than 10\% for the calculated difference between the g-factors of the upper (${\rm g}^u$) and lower (${\rm g}^l$) Stark doublet levels of $F=3/2$, which is the main topic of this paper.

\subsection{Electronic structure calculation}
Off-diagonal matrix elements (\ref{Gperp1})--(\ref{Delta6}) were calculated using the relativistic Fock-space coupled cluster singles and doubles method (FS-CCSD)~\cite{Eliav:Review:22} with the finite-order wave function expansion technique~\cite{Zaitsevskii:ThO:23,Oleynichenko:Optics:23}. Sixty inner-core electrons were excluded from the correlation treatment using the valence part of the generalized relativistic effective core potential (GRECP) method for Th~\cite{Titov:99,Skripnikov:13c,Mosyagin:16}. In these calculations, the uncontracted triple-zeta Dyall's AE3Z~\cite{Dyall:06,Dyall:12,Dyall:2016} basis set was employed for F. The basis set for Th was obtained by selecting $s, p, d, f$ primitive Gaussians from the uncontracted quadruple-zeta Dyall AE4Z basis set, supplemented by $4s3p2s3f$ diffuse functions. Gaussian basis functions with exponential parameters larger than 500, which are necessary for describing wave function oscillations in the inner-core region of Th, were excluded due to the use of the effective core potential technique. To construct the $g$, $h$, and $i$ harmonics, we employed the procedure for generating relativistic natural-like~\cite{Roos:05} compact basis sets~\cite{Skripnikov:13a,Skripnikov:2020e,Athanasakis2025,Skripnikov:2021a} (see also Ref.~\cite{Oleynichenko24102024}). In this approach, we averaged density matrices obtained at the scalar-relativistic coupled-cluster calculations with single, double, and perturbative triple excitations CCSD(T)~\cite{Bartlett:2007} level for the Th$^{2+}$ ion, considering its low-lying electronic states. Additionally, we incorporated atomic Th blocks of relativistic two-component density matrices obtained from two-component (i.e. with spin-orbit effects included) Hartree-Fock calculations of ThF$^{3+}$ and ThF$^{+}$ ions, including both occupied and low-lying virtual spinors. The inclusion of these density matrices ensures that the necessary functions to describe polarization and spin-orbit effects for the given molecule and its ions are incorporated into the basis set. In both scalar-relativistic and two-component relativistic calculations described above, we used the extended AE4Z basis set for $s, p, d, f$ harmonics, supplemented by $15g, 15h$, and $15i$-type functions taken in an even-tempered mode to provide sufficient flexibility. After diagonalizing the resulting eigenvalues and eigenvectors of the operator corresponding to the averaged density matrix, we selected $6g, 5h$, and $3i$ compact contracted basis functions with the largest eigenvalues. To reduce the computational cost of subsequent molecular calculations, the selected contracted functions were reexpanded over a reduced number of specially optimized primitive functions: $10g, 8h$, and $6i$.

The target electronic states of ThF$^+$ were obtained in the 0-hole, 2-particle sector of Fock space, with the ThF$^{3+}$ charge state chosen as the Fermi vacuum. We employed the intermediate Hamiltonian method for incomplete main model spaces~\cite{Zaitsevskii:QED:22} in the sector with two particles. The main model space was selected to describe the $7s^2$, $7s^1 6d^1$, and $7s^1 5f^1$ valence electron configurations of Th in ThF$^+$, as all excited electronic states of interest correspond to these configurations. The total number of spinors included in the model space was set to 34. The virtual orbital energy cutoff was set to 300 $E_h$ to ensure proper correlation of the outer-core electrons~\cite{Skripnikov:17a}.


Relativistic two-component Hartree-Fock calculations, as well as molecular integral transformations, were carried out using the {\sc dirac} code~\cite{DIRAC19,Saue:2020}. Fock-space coupled-cluster calculations were performed with the {\sc exp-t} code~\cite{EXPT_website,Oleynichenko_EXPT,Zaitsevskii:ThO:23,Oleynichenko:Optics:23}. All scalar-relativistic calculations were performed using the {\sc cfour} code~\cite{CFOUR}.
A natural-like relativistic compact basis set was constructed using the {\sc natbas} code~\cite{Skripnikov:13a,Skripnikov:2020e,Athanasakis2025}. Matrix elements of the electronic angular momentum operators over molecular spinors were calculated using the {\sc oneprop} code developed in Ref.~\cite{Skripnikov:13c,Skripnikov:16b,Petrov:14}.

\section{Results}

We define the g-factors such that the Zeeman shift is given by
\begin{equation} 
   E_{\rm Zeeman} = -{\rm g}\mu_B \B M_F,
 \label{Zeem}
\end{equation}

\begin{figure}  
\includegraphics[width=0.5\textwidth]{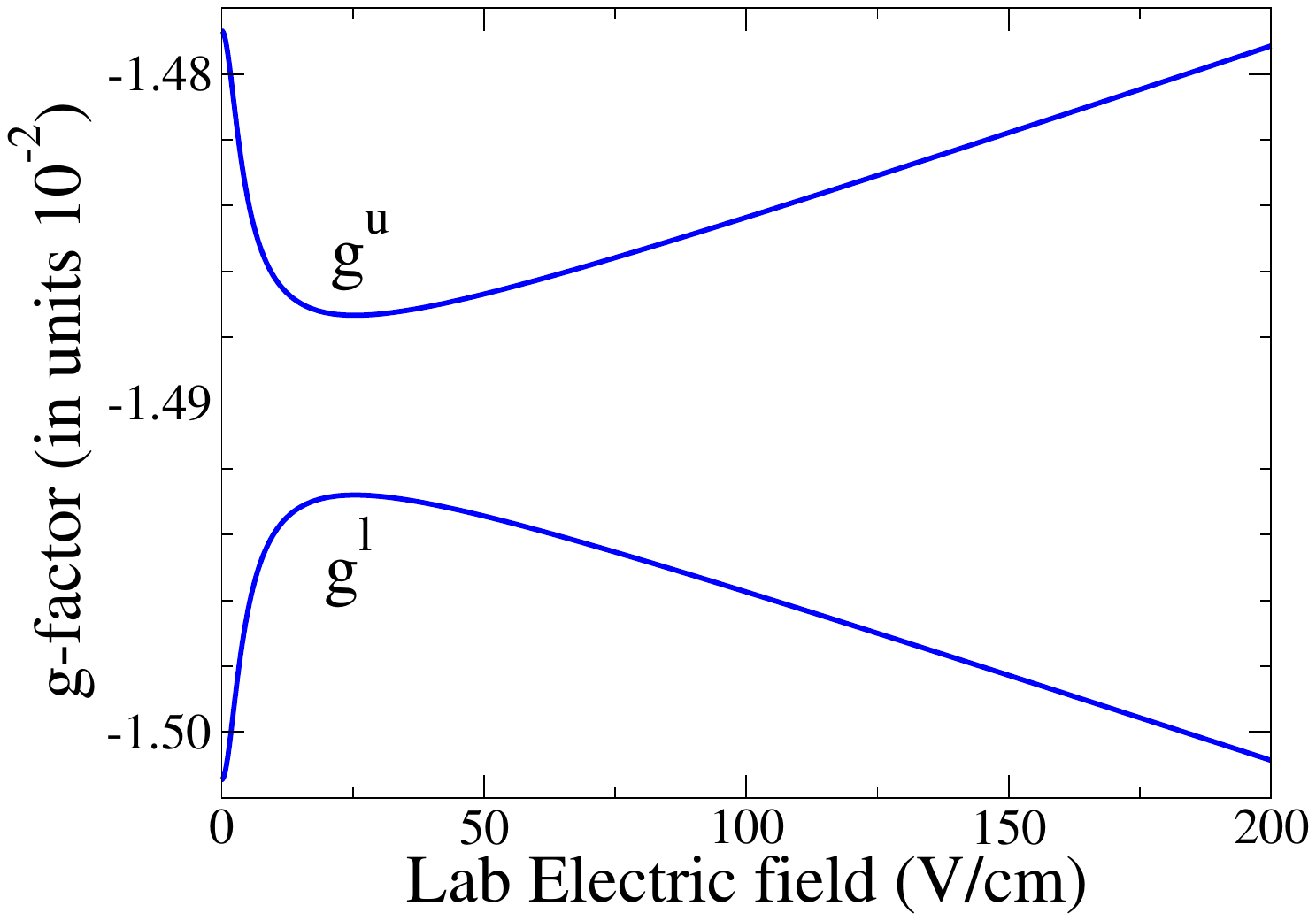}
 \caption{(Color online) Calculated g-factors for the upper (${\rm g}^u$) and lower (${\rm g}^l$) levels of the Stark doublet for the $J = 1, F=3/2, |M_F|=3/2$ levels of the $^3\Delta_1$ state of ThF$^+$ as functions of the electric field.}
 \label{gfgecross}
\end{figure}

In Fig.~\ref{gfgecross}, we plot the calculated g-factors for the $J = 1, F=3/2, M_F=3/2$ levels of the ThF$^+$ $^3\Delta_1$ state as functions of the laboratory electric field. The calculated difference $\Delta{\rm g} ={\rm g}^u - {\rm g}^l = 2.3\times10^{-4}$ between the g-factors of the upper (${\rm g}^u$) and lower (${\rm g}^l$) levels of the $\Omega$ doublets for a zero electric field is in agreement with the experimental value $|\Delta{\rm g}| = 3(3)\times10^{-4}$ \cite{Ng:2022} (the sign of $\Delta{\rm g}$ was not determined in Ref. \cite{Ng:2022}).

In the \eEDM\ experiment, the measurement of the energy split ($f$) between states with different directions of total angular momentum is repeated for different orientations of the magnetic field and for different (upper and lower) Stark doublet levels. 
%
The \eEDM\ signal manifests as a contribution to the $f^{ {\cal B} {\cal D}}$ component of the energy splitting $f$ following the relation $f^{ {\cal B} {\cal D}} = 2d_{e}E_{\rm eff}$. 
 $f^{ {\cal B} {\cal D}}$ is a component of the energy splitting $f$ which is odd under both reversals  (see Refs. \cite{Cornell:2017, newlimit1, Petrov:23b} for details). 

However, experimental equipment is never perfect; thus, beyond the \eEDM, there are multiple systematic effects contributing to $f^{{\cal {B}}{\cal {D}}}$ that can mimic the \eEDM\ signal \cite{newlimit2}. 
For instance, as shown in Ref.~\cite{newlimit2}, contributions to $f^{{\cal {B}}{\cal {D}}}$ arise from the {\it non-reversing component of the magnetic field} and the {\it ellipticity of the rotating electric field}. Additionally, as demonstrated in Ref. \cite{Petrov:23b}, the {\it second and higher harmonics of the rotating electric field} also contribute to $f^{{\cal {B}}{\cal {D}}}$. For brevity, systematic effects corresponding to the non-reversing component of the magnetic field, the ellipticity of the rotating electric field, and the second and higher harmonics of the rotating electric field will be referred below as systematic effects 1, 2, and 3 respectively. Notably, these systematic effects contribute more significantly to the $f^{\cal {B}}$ component of the splitting, which is odd with respect to the reversal of the magnetic field.

The knowledge of the ${\rm g}^u$ and ${\rm g}^l$ dependence on the electric field allows for the determination of the ratio
${f^{\cal B}}/{f^{ {\cal B} {\cal D}}}$ 
for systematic effects 1, 2, and 3. Thus, by monitoring $f^{\cal {B}}$ (which is significantly larger than $f^{{\cal {B}}{\cal {D}}}$, see \tref{deltag01}) during the experiment, one can estimate the contributions beyond \eEDM\ to $f^{{\cal {B}}{\cal {D}}}$ and thereby establish a method for controlling and accounting for systematic effects \cite{newlimit2}.
According to Refs. \cite{newlimit2, Petrov:23b}, for systematic effect 1, under a good approximation, we have
 \begin{equation}
S_1 = \frac{f^{\cal B}}{f^{ {\cal B} {\cal D}}} = \left( \frac{ {\rm g}^u +{\rm g}^l} { { \rm g}^u -{\rm g}^l }  \right).
 \label{ratio}
\end{equation}
 Within a small area of the value of static electric field, the g factor difference can be presented as
    \begin{equation}
    \Delta {\rm g} = \Delta {\rm g}_0 +  \Delta {\rm g}_1 E.
    \label{ratio2}
    \end{equation}
Then according to Ref. \cite{Petrov:23b} we have
    \begin{equation}
    S_2 =  \frac{f^{\cal B}}{f^{ {\cal B} {\cal D}}} = \left( \frac{\Delta {\rm g}_0} {\Delta {\rm g}} + \frac{3}{4} \frac{\Delta {\rm g}_1 E} {\Delta {\rm g}} \right) \left( \frac{ {\rm g}^u +{\rm g}^l} { { \rm g}^u -{\rm g}^l }  \right)
    \end{equation}
    for systematic effect 2 and 
  \begin{equation}
S_3 = \frac{f^{\cal B}}{f^{ {\cal B} {\cal D}}} = \left( \frac{ {\rm g}^u +{\rm g}^l} { \Delta {\rm g}_0 }  \right).
 \label{ratio3}
\end{equation}
    for systematic effect 3.
    
In \tref{deltag01}, the calculated $\Delta {\rm g}_0$, $\Delta {\rm g}_1$, and ratios ${f^{\cal B}}/{f^{ {\cal B} {\cal D}}}$ for systematic effects 1, 2, and 3 are presented. The ratios $S_1$, $S_2$, and $S_3$ differ, making it more challenging to control systematic effects, as it is generally unknown which specific systematic effect (or combination thereof) caused the nonzero values of $f^{{\cal {B}}{\cal {D}}}$ and $f^{\cal {B}}$ \cite{newlimit2}. 
For HfF$^+$ at an electric field of $E=58$~V/cm, we have $S_2 = 0.76 S_1$ and $S_3 = 34 S_1$ \cite{Petrov:23b}. According to \tref{deltag01}, for ThF$^+$ at an electric field of $E=60$ V/cm, one can obtain $S_2 = 0.83 S_1$ and $S_3 = 3.2 S_1$. The values of $S_i$ (especially for $S_3$) are closer to each other for ThF$^+$ than for HfF$^+$, which provides an advantage for ThF$^+$.

\begin{table}
\caption{The calculated $\Delta{\rm g}_0$, $\Delta {\rm g}_1$ (cm/V), and ratios ${f^{\cal B}}/{f^{ {\cal B} {\cal D}}}$ for systematic effects 1, 2, and 3 ($S_1$, $S_2$, and $S_3$, respectively).}
\label{deltag01}
\begin{tabular}{cccccc}
\hline
\hline
E(V/cm)                     &  $10^7 \Delta{\rm g}_0$ & $10^7\Delta {\rm g}_1$ & $S_1$ & $S_2$ & $S_3$ \\
\hline
      40      &      349.4     &         6.3       &          -494     &       -442     &       -853   \\
      50      &      280.2     &         7.9       &          -442     &       -377     &      -1064   \\
      60      &      233.8     &         8.7       &          -393     &       -325     &      -1275   \\
      70      &      200.6     &         9.3       &          -351     &       -284     &      -1486   \\
      80      &      175.6     &         9.6       &          -316     &       -252     &      -1697   \\
      90      &      156.1     &         9.8       &          -287     &       -226     &      -1909   \\
     100      &      140.6     &        10.0       &          -262     &       -204     &      -2120   \\
     110      &      127.8     &        10.1       &          -241     &       -187     &      -2331   \\
     120      &      117.2     &        10.2       &          -222     &       -172     &      -2543   \\
     130      &      108.2     &        10.3       &          -207     &       -159     &      -2754   \\
     140      &      100.5     &        10.3       &          -193     &       -148     &      -2966   \\
     150      &       93.8     &        10.4       &          -181     &       -138     &      -3177   \\
\hline
\hline     
\end{tabular}
\end{table}

\section{Conclusion}
We have considered the g-factors of the $J = 1$, $F=3/2$, $|M_F|=3/2$ hyperfine levels of the \eEDM-sensitive ground electronic state $^3\Delta_1$ of the $^{232}$ThF$^+$ cation in an external electric field. The calculated difference between the g-factors of the upper and lower levels of the $\Omega$ doublets is in agreement with the recent experiment within the experimental error.
Systematic effects related to the non-reversing component of the magnetic field, the ellipticity of the rotating electric field, and the second and higher harmonics of the rotating electric field are considered.

Our
results for the g-factors and their difference in the ground electronic state of ThF$^+$ provide key input parameters for an upcoming generation of \eEDM~experiments. By offering a quantitative handle on the dependence of the g factors upon external electric fields, this work paves the way for enhanced control of magnetic-field-related systematic effects. We anticipate that our findings will be instrumental in interpreting future measurements of ThF$^+$.

\section{Acknowledgements}
Calculations of the g factor difference for $^{232}$ThF$^+$ are supported by the Russian Science Foundation grant no. 24-12-00092 (\url{https://rscf.ru/project/24-12-00092/}).
Electronic calculations of the molecular terms were supported by the Foundation for the Advancement of Theoretical Physics and Mathematics ‘BASIS’ Grant according to Project No. 24-1-1-36-1.



\end{document}